\documentclass[12pt,preprint]{aastex}

\shorttitle{Magnetic Field in Galactic Center}
\shortauthors{}
\begin{document}
\input psfig.sty
\title{Turbulent Origin of the Galactic-Center Magnetic Field: Nonthermal Radio Filaments}
\author{Stanislav Boldyrev\altaffilmark{1} and Farhad Yusef-Zadeh\altaffilmark{2}}
\altaffiltext{1}{Department of Astronomy and Astrophysics, 
University of Chicago, 5640 S. Ellis Ave., Chicago, IL 60637, {\sf boldyrev@uchicago.edu}}
\altaffiltext{2}{Department of Physics and Astronomy, Northwestern University, 
Evanston, IL 60208, {\sf zadeh@northwestern.edu}}

\input psfig.sty

\begin{abstract} 

A great deal of study has been carried out over the last twenty years 
on the origin of the magnetic activity in the Galactic center. One of the 
most popular hypotheses assumes  
milli-Gauss magnetic field with poloidal geometry, pervading the 
inner few hundred parsecs of the Galactic-center region. However, 
there is a growing observational evidence for the large-scale 
distribution of a much weaker field of $B \lesssim 10 {\rm \mu}G$  in this 
region. Here, we propose that the Galactic-center magnetic field 
originates from turbulent activity that is known to be extreme 
in the central hundred parsecs. In this picture the spatial 
distribution of the magnetic field energy is highly intermittent, and 
the regions of strong field have filamentary structures. We propose 
that the observed  nonthermal radio filaments appear in (or, possibly, 
may be identified with) such strongly 
magnetized regions. At the same time, the large-scale diffuse magnetic field is weak.  
Both results of our model can 
explain the magnetic field measurements of the 
the Galactic-center region. In addition, we discuss 
the role of ionized outflow from stellar clusters  in  producing  
the long magnetized filaments perpendicular to the Galactic plane. 

\end{abstract}
\section{Introduction.}
\label{introduction} 

Magnetic field is ubiquitously inferred in the interstellar medium. 
Although its origin and large-scale (Galactic-scale) structure is still 
debated, it is suggested by observations that the Galactic field strength 
is of the 
order of few $\mu G$~\citep[e.g.,][]{zweibel}. Moreover, the fluctuating 
part of the field is of the order of its large-scale 
component, and that the field is roughly in equipartition with the gas
pressure and the cosmic-ray pressure. 

Less consensus has been reached about the origin, structure, and 
strength of the 
magnetic field in the center of the Galaxy (central hundreds of parsecs). 
The earliest study of large-scale  magnetic activity in this region dates 
back to 
the discovery of 
nonthermal radio filaments more than 20 years 
ago~\citep{yusef-zadeh84}. 
These filaments are fairly straight, about ten to hundred parsec long 
and about a fraction of a parsec wide. The linear polarization of these 
filaments indicates that they  radiate 
due to synchrotron emission of energetic, non-thermal particles following 
the magnetic field lines (the origin of these particles being a separate 
issue).  The rigidity  and the linearity of the filaments 
distributed within the zone of molecular gas  having  high 
density and velocity dispersion 
were used to make an estimate of the 
magnetic field. 
This dynamical  argument
 estimated a milli-Gauss 
field~\citep{yusef-zadeh87,serabyn-gusten,morris96,bicknell}.  In the 
intervening years, a number of additional filaments have been discovered 
\citep[see the review 
by][]{yusef-zadeh03,lang,reich03,nord,larosa04,yusef-zadeh04}.

In early works, nonthermal radio filaments were detected predominantly 
perpendicular to 
the Galactic plane. This prompted the theories suggesting 
that strong magnetic field, of the order of milli-Gauss, is uniformly 
pervasive in the Galactic center, and relativistic electrons are injected into 
this pre-existing field at some places thus ``lighting up''  the 
corresponding   
magnetic-field lines. One of such theories  assumed  
that the magnetic flux was already strong enough 
in the protogalaxy, and then it was confined to the Galactic disk 
during the Galactic disk formation. The subsequent matter inflow 
into the center further concentrated the poloidal 
flux~\citep[see e.g.,][]{sofue,benford88,morris94,chandran}.  
Alternatively, certain {\em local} mechanisms have been proposed to explain the 
production of non-thermal radio filaments \citep[see e.g.,][]{rosner96,shore99,yusef-zadeh03}. 
In one model, the magnetic field was amplified by thermal 
instability \citep{rosner96}, whereas in another the 
filaments were considered to be magnetized wakes produced as a result of an 
interaction of the molecular clouds with a Galactic center 
wind~\citep{chevalier92,shore99}. 

The conventional theory that assumes milli-Gauss magnetic field with 
poloidal geometry, surrounding the Galactic center, 
needs to explain not only the origin of $\sim 4\times 10^{54}$ ergs 
of energy trapped in the magnetic field within a 100~pc radius,  
but also the mechanism of anchoring of the strong magnetic field lines.  
Moreover, this theory has recently encountered a number of serious observational 
challenges. First, more than 
eighty radio filaments have already been detected, and their directions do 
not always run perpendicularly to the galactic 
plane~\citep[e.g.,][]{lang,larosa04,nord,yusef-zadeh04}.  Second, a number of arguments 
based on Zeeman and Faraday-rotation measurements toward the Galactic 
center as well as on the estimates of synchrotron life-time, suggest that 
the pervasive magnetic field strength should not exceed $100 {\mu}$G 
\citep{uchida95,yusef-zadeh03,larosa05}.  Third, the anisotropic 
property of the 
scattering medium, which is thought to be seated  in the Galactic center 
region,  
shows a random orientation of scatter-broadened OH/IR stars in the central 
50$'$ of the Galactic center \citep{vanlangevelde92,frail}.  The combined 
anisotropy of OH/IR stars, OH (1720 MHz) masers, as well as the anisotropy 
of the rotation measure of the nonthermal filament G359.1-0.54 
\citep{yusef-zadeh97} are inconsistent with a strong poloidal 
space-filling 
magnetic field 
pervading the Galactic-center region. Fourth, the recent discovery of X-ray 
emission from three nonthermal filaments \citep{lu03,sakano} has suggested 
that X-ray emission from G359.90-0.06 could be 
produced by inverse Compton scattering (ICS) of far-infrared photons from 
dust by the relativistic electrons responsible for the radio synchrotron 
emission \citep{yusef-zadeh05}. The production of X-ray emission from ICS 
allows an estimate of the magnetic field strength of order ~0.1 mG within 
the 
nonthermal filament.  Fifth, sub-millimeter  polarization measurements of 
dust emission from several Galactic-center molecular clouds show that  
the large-scale distribution of the magnetic field follows a 
toroidal geometry along the 
Galactic plane \citep{novak00,chuss03}. And last, the 
recent discovery of diffuse nonthermal 
structure in the Galactic center by \citet{larosa05} imposes even stronger 
limitation of the large-scale magnetic field strength, suggesting that the 
pervasive magnetic field in the Galactic center is rather {\em weak}, of 
order~$10\mu G$. This is consistent with earlier measurements of 
rotation measure 
distribution of extragalactic radio sources projected toward the 
Galactic center \citep{roy}, and with the observations of the nonthermal 
filament G359.1-0.2 \citep{gray95},  indicating 
magnetic field strength of the Faraday screen, ranging between 1 
and 10 $\mu$G.

In other words, several observations indicate that although the magnetic 
field in 
the non-thermal filaments may be strong, it is not representative of the 
large-scale pervasive magnetic field in the Galactic  center. 
A theory of the Galactic center is thus in demand, which would 
provide a robust explanation for both weak pervasive magnetic field and 
local, randomly  oriented, strongly magnetized filamentary structures. 
In the present paper we attempt to propose such a unifying explanation. We 
suggest that, quite generally, magnetic structure of the Galactic center 
can be derived from its strong turbulent activity. We argue that the 
pervasive magnetic flux in the Galactic center is quite weak, however, the 
magnetic field is significantly amplified {\em locally} so that the regions of strong 
field have filamentary structures, as an inevitable 
consequence of magnetic dynamo mechanism associated with such a turbulent 
activity. We then propose that the nonthermal radio filaments are observed 
in (and possibly may be identified with) such strongly magnetized regions. 
A physical picture of strongly turbulent Galactic center is 
motivated by radio observations of compact sources indicating heavily 
scatter-broadened radio sources toward the inner degree of the Galactic 
center \citep{vanlangevelde92}.  The most extreme scattering medium in the 
Galaxy is argued to be located within the inner few hundred parsecs of the 
Galactic center \citep{yusef-zadeh94,lazio98}. This  scattering medium is 
considered to be denser than in the disk, $n_e\sim 1$~cm$^{-3}$, by one to 
two orders of magnitude \citep{lazio98}.

\section{Inhomogeneous MHD Turbulence} 
\label{turbulence}
Magnetohydrodynamic turbulence is generally investigated, both 
analytically and numerically, for idealized homogeneous and isotropic 
settings. However, astrophysical turbulence is almost always 
inhomogeneous and anisotropic. In the situation when the outer scale of 
turbulence (the correlation scale of turbulent fluctuations) is 
much smaller than the scale of turbulent intensity variation, 
the approximation of homogeneity works well. However, as we show 
in the next section, in the Galactic center these two scale lengths  
are of the same order. In this case, the effect of magnetic-field amplification 
by turbulent motion should be considered in conjunction with another effect --
magnetic-field expulsion from the region of stronger turbulence. In the present 
section we propose a description of magnetic-field structure in inhomogeneous  
turbulence.  

As first noted by 
\citet{batchelor}, when fluid viscosity and resistivity are negligible, and 
if the magnetic field acting back on the fluid is weak, 
the evolution equation for magnetic field, $\partial_t {\bf B}=\nabla \times \left[{\bf v} 
\times {\bf B} \right]$,  
formally coincides 
with the equation for the vorticity, ${\bf \omega}=\nabla \times {\bf v}$.  
When a weak diffuse magnetic field is amplified 
by turbulence, it is randomly stretched and folded so that magnetic energy 
distribution becomes spatially intermittent. Analogously  to a distribution 
of vorticity in hydrodynamic turbulence \citep[e.g.,][]{she,frisch,kaneda}, 
the strongest field is concentrated 
in filamentary structures, where it reaches the 
equipartition with the turbulent energy \citep[e.g.,][]{nordlund}. 
The characteristic rate of field amplification is the turbulent-eddy 
turnover rate, and the characteristic length of the filamentary structures 
is the outer-scale of turbulence.

If the magnetic field were confined to the turbulent region (as, for 
example, in most numerical simulations), the number of filamentary structures 
would increase until the magnetic field would become strong 
everywhere~\citep[e.g.,][]{cattaneo}. If, however, turbulence is 
confined to some spatial region but magnetic field is not, 
then the so-called diamagnetic effect comes into play. Namely, quite 
generally, large-scale magnetic flux tends to be expelled from the turbulent 
region. In other worlds, when the intensity of fluid turbulence varies 
in space, the large-scale magnetic field is transported in the direction of weaker 
turbulence~\citep{zeldovich,parker75,vainshtein-kichatinov,landau}.  
The characteristic time for flux expulsion  is the turbulent diffusion 
time. Denote $l_0$ the largest scale of turbulence, and $v_0$ the eddy turnover 
velocity at this scale, then this time is estimated as $\tau_d\sim L^2/\eta_T$, where $L$ is the size of 
the turbulent region (we assume $l_0<L$) and $\eta_T=l_0v_0$ is 
turbulent diffusivity.
As we discuss in the next section,    
the size of the turbulent region in the Galactic center is comparable 
with the outer scale of turbulence, or may be bigger by a factor of few, 
$L\gtrsim l_0$. 
We may therefore conjecture that in a steady state 
the magnetized filamentary structures have a chance to diffuse out of the turbulent 
region, so that the magnetic field does not become strong everywhere in the turbulent 
region. Rather, the strongest field is embedded by magnetic flux tubes.
 
We propose that these magnetic structures are the places where the 
Galactic-center non-thermal radio filaments appear. In this model relativistic 
particles may be pervasive (say, cosmic rays), while the regions of strong 
magnetic field are spatially intermittent, and the average magnetic field is 
weak. The synchrotron emissivity that is proportional to $B^{(1+\gamma)/2}$, 
where $\gamma$ is the spectral index of the power-law energy spectrum of 
relativistic particles \citep{salter-brown} and $B$ is the component of the 
magnetic field perpendicular to the line of sight, is enhanced along the 
filaments as the magnetic field is amplified. For example, in the case of a 
steep spectrum of relativistic particles measured at high frequencies, with 
$\gamma\sim 3$ \citep{yusef-zadeh05,anantha91,lang} the synchrotron emission 
may increase by two orders of magnitude with respect to background 
emission if the magnetic field is amplified by only a factor of~10, assuming 
that the cosmic-ray particle density is constant. In the case of 
low value of $\gamma\sim 2$, as  measured at low frequencies 
~\citep{larosa00}, the   synchrotron emissivity could increase only by  a 
factor of 30. 
Alternatively, however, the 
relativistic particle density may not be constant. For example, these particles 
could be injected locally by compact sources that produce relativistic 
particles \citep{yusef-zadeh-konigl}. We now apply the above general arguments 
to estimate the parameters of the nonthermal radio filaments.

\section{Magnetic-Field Structure in the Galactic Center} 
\label{dynamo} 
We consider  that characteristic gas distribution in the 
Galactic center 
consists of four  phases: cold, cool,  warm, and hot \citep{oka}. 
The cold phase is a molecular gas 
whose temperature is $T_{cold}\sim70{\rm K}$, and density 
is $n_{cold}>10^4{\rm cm}^{-3}$. These molecular clouds occupy only 
a small fraction of space, with a filling factor~$f_{cold}\lesssim 0.01$. 
The cool phase consists of  atomic hydrogen with temperature $T_{cool}\sim 
250{\rm K}$, 
and density $n_{cool}\sim 10^2{\rm cm}^{-3}$. Its filling factor can 
be large, $f_{cool}\sim 0.5-1$. The worm phase is ionized gas, with 
temperature $T_{warm}\sim 10^4{\rm K}$, 
density $n_{warm}\sim 10^2{\rm cm}^{-3}$ and filling 
factor $f_{warm}\lesssim 1$ based on the scattering measure  of the Galactic 
center sources \citep{lazio98}.  The hot phase of the Galactic 
center has been modeled to have two plasma components,  the soft component  
with kT~$\sim 0.8$~keV and the hard one with a kT~$\sim$~8~keV 
\citep[e.g.,][]{muno,koyama96}. The mean electron  density of hot gas 
is  estimated to be $n_{hot}\sim$~0.3--0.4cm$^{-3}$, and the 
filling factor is~$f_{hot}\lesssim 1$ \citep{koyama96}.  
Considering that all these phases co-exist with each other, 
 it  is difficult  to argue that they are in thermal 
pressure equilibrium. Rather, observations of  molecular and 
atomic line emission \citep[e.g.,][]{tsuboi99,oka,geballe05} suggest that 
these 
phases are turbulent with turbulent velocity dispersion 
of order $v_0\sim 10 - 50 {\rm km/sec}$.  We therefore believe 
that it is 
reasonable to assume  that their turbulent energies rather than thermal 
energies are of the same order, and that strong turbulence plays an essential role 
in the Galactic-center gas dynamics. 

To estimate the outer scale of the turbulence, we need to specify possible 
mechanisms of turbulence generation. We may first assume that the 
turbulence is stirred through supernovae explosions, and accept that the 
rate of explosions in the central 50 pc is about 1 per $10^5$ 
years~\citep{figer,larosa05}. Assuming that a supernova shell expands with velocity~$v_{sh}\sim 10^7
{\rm cm/s}$, we deduce that the outer scale of 
turbulence in the Galactic center environment is of 
order $l_0\sim 30 {\rm pc}$. (A shell expands uninterrupted by another supernovae 
explosion up to a radius $l_0$ during time $l_0/v_{sh}$. Another supernova will distort 
the expanding shell if the following condition is satisfied, 
$l_0^3/[50 {\rm pc}]^3\sim [v_{sh}/l_0]10^5 {\rm yrs}$, which leads to the above 
magnitude of $l_0$).  This value provides a simple, 
but quite rough estimate.  A more plausible consideration is that the large-scale  
turbulence is not uniform and isotropic. Stronger turbulence may be produced by massive 
star-forming activity in the Galactic center, which appears to occur predominantly 
in star clusters. 
In this picture turbulence could be driven by strong stellar 
winds from these young massive clusters.  
These winds can collide  
on scales characterized by  the distance between 
mass-losing stars within a cluster or clusters themselves, $\sim 
10-20 {\rm pc}$ along  the Galactic plane. The winds could also 
be correlated on much larger 
scales $\sim 100{\rm pc}$ perpendicular to the Galactic plane (where the gas 
density drops off),   
driving thermal gas 
away in this direction.  

A number of studies have recently suggested that the high pressure environment 
of the Galactic center went through a mini-starburst activity ten million years 
ago \citep[e.g.,][]{bland-hawthorn}. In particular, the large scale 1$^0$ 
omega-shaped lobe structure \citep{sofue-handa} shows the largest 
concentration of nonthermal filaments in the Galactic center region.  This 
large concentration of magnetic filaments appears in the vicinity of the 
footprints of the large-scale lobe where the bright HII complex of Sgr C (e.g., 
G359.5-0.0) \citep{yusef-zadeh04, nord} as well as the radio continuum arc 
(l$\sim0.2^0$) are located. 
We suggest that the presence of the longest and brightest vertical 
nonthermal filaments is phenomenologically tied to the origin of the 
omega-shaped lobe structure. In this picture, the outer scale of the turbulence 
stretching the field lines is in the direction away from the plane and is 
identified by the past activity of the collision of the winds and supernovae.  
In particular, an anisotropic distribution of thermal and nonthermal gas in the 
direction along and away from the Galactic plane can naturally explain the fact 
that the longest filaments run predominantly perpendicularly to the Galactic 
plane. In this picture, we speculate that the outer scale of turbulence could 
be an order of 
magnitude larger away from the plane than parallel to the plane.  
 
When strong turbulence is switched on in a highly conducting 
medium permeated by weak, diffuse magnetic field, 
the field is amplified by random fluid motion, under the dynamo 
mechanism~\citep[e.g.,][]{kazantsev,moffatt,vainshtein-kichatinov,kulsrud-anderson,schekochihin2}. 
As we discussed in the previous section,  at the initial stage of 
amplification, the strong-field regions have morphology of 
filaments \citep[e.g.,][]{nordlund}. The magnetic field 
inside these filamentary structures increases 
until the Lorentz force is strong enough 
to prevent further field amplification.   
This happens in 
several dynamical, eddy turn-over times, 
$\tau\sim l_0/v_{0}\sim 10^6-10^7{\rm yrs}$.  
The balance of the Lorentz force inside 
a filamentary structure  
and the surrounding turbulent forces gives the magnetic field inside 
the structure, $B_f\gtrsim 0.1$~mG. 
The length 
of a filamentary structure cannot exceed the outer scale of turbulence~$l_0$, since 
the strongest eddy that can 
stretch the structure has the scale~$l_0$. The characteristic distance between 
filamentary structures is also of the order of $l_0$. However, shorter and weaker 
structures are possible; they can appear closer to each other since they are 
created by smaller and less energetic eddies.  

Assume now, that mean magnetic field in the Galactic center is roughly of the same 
order as in the rest of the Galaxy,~$B_0\lesssim 10\mu G$. The field is amplified 
by turbulent motion locally. Consider the regions of space where the field is amplified 
by at least a factor of 10 (i.e., $B_f>0.1{\rm mG}$).  Such  
strongly magnetized filamentary structures should be encountered 
with a rather small probability, i.e. they should have a rather small volume 
filling factor. To illustrate this, one can use the probability density function 
(PDF) of magnetic-field strength. Since 
the pervasive field $B_0$ is weaker than the value dictated by 
equipartition with turbulent energy, this function can be obtained from 
the kinematic dynamo theory. In the well investigated homogeneous and isotropic case, 
this theory predicts that such a function is log-normal,  
$P(B){\rm d}B =\sqrt{C/\pi} \exp\left[-C\log^2(B/B_0)\right]{\rm d}B/B, $ 
where~$C$ is some coefficient, $C\sim l_0/(v_0 t)$. 
This prediction is in good agreement with numerical simulations,~\citep[e.g.,][]{cattaneo,schekochihin}. 
If the field-amplification time is 
approximately the eddy turnover time, then  $C\sim 1$ (this crude estimate suffices 
for our illustrative purposes).    Using this formula one 
can estimate that the probability (or, the volume filling factor) 
of filaments with $B>10 B_0$ is quite small, $f\sim 10^{-4}$.  
Assuming that the number of the strongest filaments per outer scale 
of turbulence is of order one, and the length of such 
filaments is~$l_0$,  we estimate $l_0 l_f^2\sim fl_0^3$. We then find the width of 
the filaments as $l_f<0.01 l_0$. This 
rough estimate is in agreement with observations \citep{yusef-zadeh04}.   

Interestingly, the theory of MHD turbulence may also allow us to explain 
another peculiar and puzzling property of nonthermal radio filaments. Quite often, these 
filaments are observed to be split into multiple 
filaments running parallel to each other. In fact, there are more such multiplets 
than single isolated filaments \cite[see, e.g.,][]{lang,yusef-zadeh04}. We may 
speculate that such multiple  
filamentary structure could be identified with folded structure of magnetic field 
inside strongly magnetized regions, as is consistently observed in 
numerics  \citep[see, e.g.,][]{cattaneo, schekochihin}. 

In conclusion, we have proposed a new interpretation of the magnetic-field 
structure in the Galactic-center region where numerous unique nonthermal filaments 
have been detected. The main idea of our approach is that the magnetic-field 
distribution in the Galactic-center environment is inherently and universally 
related to the strong turbulent activity observed in this region. Such turbulence 
locally amplifies the diffuse background magnetic field and concentrates the strong field  
in thin and elongated, randomly oriented filamentary structures. The distribution of 
the longest filamentary structures can be anisotropic, however, due to anisotropic 
character of the large-scale turbulence driven by collisions of stellar 
winds. The nonthermal radio filaments originate inside such structures, 
or can possibly be identified with these structures themselves. 
The proposed turbulent 
origin of nonthermal filaments in the Galactic center distinguishes our approach 
from other theories in that our model is mainly driven by observations, 
it is based on quite general results of the theory of turbulence, and 
it resorts to minimal non-standard assumptions. 

We are grateful to Samuel Vainshtein and Ellen Zweibel for useful discussions. The work
of S.B. was supported by the NSF Center for Magnetic Self-Organization
in Laboratory and Astrophysical Plasmas at the University of
Chicago.

\end {document}